\documentstyle[12pt]{article}
\input amssym.def       
\input amssym.tex       

\def\goth{\frak}          
\def\double{\Bbb}

\def\cc{{\double C}}     
       
\def\rr{{\double R}}     
\def\zz{{\double Z}} 
     
\def\qq{{\double Q}}

\def\hhh{{{\double H}}}   

\def\mm{{{\cal M}}}

\def\ot{\otimes}
\def\op{\oplus}

\def\bb{\begin{eqnarray}}
\def\ee{\end{eqnarray}}
\def\eee{\nonumber\end{eqnarray}}

\newtheorem{lemma}{Lemma}[section]
\newtheorem{satz}{Theorem}[section]
\newtheorem{proposition}{Proposition}[section]

\begin{document}

\font\twelve=cmbx10 at 13pt
\font\eightrm=cmr8
\def\petit{\def\rm{\fam0\eightrm}}
\baselineskip 18pt

\begin{titlepage}
\title{On the Semi-Classical Vacuum Structure of the Electroweak Interaction}
\author{J\"urgen Tolksdorf\thanks{email: tolkdorf@euler.math.uni-mannheim.de}\\
Inst. of Mathematics\\ University of Mannheim, Germany}
\date{Oct. 17, 2003}
\maketitle

\begin{abstract}
It is shown that in the semi-classical approximation of the electroweak sector 
of the Standard Model the moduli space of vacua can be identified with the first 
de Rham cohomology group of space-time. This gives a slightly different physical 
interpretation of the occurrence of the well-known Ahoronov-Bohm effect. 
Moreover, when charge conjugation is taken into account, the existence of a
non-trivial ground state of the Higgs boson is shown to be equivalent to the 
triviality of the electroweak gauge bundle. As a consequence, the gauge bundle 
of the electromagnetic interaction must also be trivial. Though derived at 
``tree level'' the results presented here may also have some consequences for 
quantizing, e. g., electromagnetism on an arbitrary curved space-time. 
\end{abstract}
\end{titlepage}

\section{Introduction}
We consider as a specific Yang-Mills-Higgs (YMH) gauge theory the bosonic part
of the electroweak interaction of the Standard Model of particle physics. Its
basic geometrical objects are given by a ${\rm SU(2)}\times{\rm U(1)}$ gauge
potential $A=W+B\in\Omega^1(\mm,\rr^{\mbox{\tiny 3}}\op\rr)$ together 
with a complex vector field $\Phi\in\Omega^0(\mm,\cc^{\mbox{\tiny 2}})$.
Here, $\mm$ denotes a space-time manifold which is usually identified with 
Minkowski space $\rr^{\mbox{\tiny 1,3}}$.  Like in perturbation theory, the 
physical interpretation of the pair $(A,\Phi)$ is that of a ``fluctuation of the 
(classical) bosonic vacuum'' $(A=0,\Phi=0)$ via the known replacement 
\bb
\label{mincoup}
d&\mapsto& d_{\!\mbox{\tiny A}}:= d + A,\cr 
{\bf z}_0&\mapsto&\phi:={\bf z}_0 + \Phi,
\ee 
where ${\bf z}_0\in\cc^{\mbox{\tiny 2}}$ is a chosen minimum of the 
Higgs potential ${\rm V}_{\!\mbox{\tiny H}}:=\lambda|{\bf z}|^{\mbox{\tiny 4}}-
\mu^{\mbox{\tiny 2}}|{\bf z}|^{\mbox{\tiny 2}}\;(\lambda,\mu>0)$ and ``$d$'' is the 
exterior derivative.\\ 

From a geometrical point of view, a (classical) bosonic vacuum of the 
electroweak interaction may be represented by the canonical YMH pair 
\bb
\label{symhpair}
(\Theta_{\!\mbox{\tiny 0}},{\cal V}_{\!\mbox{\tiny 0}}),
\ee 
where  ${\cal V}_{\!\mbox{\tiny 0}}$ is the canonical mapping
\bb
{\cal V}_{\!\mbox{\tiny 0}}:\,\mm&\longrightarrow&
\mm\times{\rm orbit}({\bf z}_0)\cr
x&\mapsto&(x,{\bf z}_0)
\ee
and $\Theta_{\!\mbox{\tiny 0}}$ is the flat connection
on ${\rm pr}_1:\,\mm\times\cc^2\rightarrow\mm$ associated with the canonical 
connection ${\rm pr}_2^*\zeta^{\mbox{\tiny MC}}$ on the trivial principal 
${\rm SU(2)}\times{\rm U(1)}$ bundle
\bb
{\rm pr}_1:\,\mm\times({\rm SU(2)}\times{\rm U(1)})&\longrightarrow&\mm\cr
p=(x,g\equiv(g_{\mbox{\tiny (2)}},g_{\mbox{\tiny (1)}}))&\mapsto&x.
\ee 
Here respectively, $\zeta^{\mbox{\tiny MC}}$ is the Maurer-Cartan form on
${\rm SU(2)}\times{\rm U(1)},$ ${\rm pr}_2(x,g):=g$ and
${\rm orbit}({\bf z}_0)\subset\cc^{\mbox{\tiny 2}}$ denotes the orbit of ${\bf z}_0$ 
with respect to the unitary representation $\rho_{\mbox{\tiny H}}(g):=
g_{\mbox{\tiny (2)}}\,g_{\mbox{\tiny (1)}}^{\mbox{\tiny y}}\;({\rm y}\in\qq)$. 
Notice that the specific Higgs potential ${\rm V}_{\!\mbox{\tiny H}}$ in 
the Standard Model has only one such orbit that is homeomorphic to 
${\rm S}^{\mbox{\tiny 3}}\subset\rr^{\mbox{\tiny 4}}$.\\

The canonical YMH pair (\ref{symhpair}) represents a specific absolute minimum 
of the energy functional associated with the known YMH action 
${\cal I}_{\!\mbox{\tiny YMH}} = {\cal I}_{\!\mbox{\tiny YM}} +
{\cal I}_{\!\mbox{\tiny H}}$. 
Of course, any gauge equivalent YMH pair contains the same physical information. 
A question that naturally follows is how many gauge inequivalent vacua of the 
electroweak interaction exist and what is their physical meaning? 
Another question closely tied to the previous one is: How do we know that 
the gauge bundle underlying the electroweak interaction is actually trivial? In 
other words, what can we learn from the study of the moduli space of vacua 
concerning the topology of space-time and the gauge bundle? At a
first glance this question may sound like being of purely mathematical interest. 
However, the existence of gauge inequivalent ground states, and tied to it the 
topology of space-time and of the gauge bundle, may also have consequences 
with respect to the quantization of a spontaneously broken gauge theory. For
instance, when trying to quantize electromagnetism on an arbitrary
(globally hyperbolic) space-time manifold $\mm$ one has to consider the 
non-triviality of ${\rm H}^2_{\mbox{\tiny deR}}(\mm)$. However, when seen
from a gauge geometrical viewpoint, the Maxwell-Faraday equation 
$dF_{\!\mbox{\tiny elm}}=0$ becomes an identity (the ``Bianchi Identity''). That 
is, the electromagnetic field strength $F_{\!\mbox{\tiny elm}}\in\Omega^2(\mm)$ 
is considered as the local pull-back of the curvature of a connection form 
$\omega\in\Omega^1({\rm Q})$ on the underlying electromagnetic gauge
bundle ${\cal Q}$: $\pi_{\mbox{\tiny Q}}:\,{\rm Q}\rightarrow\mm$. In other
words, $F_{\!\mbox{\tiny elm}}=d\sigma_{\!\!\mbox{\tiny$\alpha$}}^*\omega,$ 
where $\sigma_{\!\!\mbox{\tiny$\alpha$}}:\,
\mm\supset{\rm U}_{\!\mbox{\tiny$\alpha$}}\rightarrow{\rm Q}$ is a local
trivialization of ${\cal Q}$. Therefore, in order to quantize the electromagnetic
gauge potentials 
$A_{\mbox{\tiny$\alpha$}}\equiv\sigma_{\!\!\mbox{\tiny$\alpha$}}^*\omega$ 
one not only has to take into account the topology of space-time $\mm$ but, in 
particular, the topology of the electromagnetic gauge bundle ${\cal Q}$. In fact, 
if the latter turns out to be trivial, then every gauge potential 
$A\in\Omega^1(\mm)$ is a globally defined object independent of the topology 
of space-time.\\

Thus, also from a physical perspective it seems appropriate to put the 
above geometrical interpretation of a vacuum (or its ``fluctuation'') in a 
more general geometrical perspective. A corresponding discussion of a general 
(classical) bosonic vacuum can be found in \cite{tolk1'03} (for a discussion of the
fermionic vacuum, please see \cite{tolk2'03}). There, we also discussed the
geometrical meaning of the bosonic (resp. the fermionic) ``mass matrix'' and
the existence of the ``unitary gauge''. For the convenience of the reader we
shall summarize in the next section the basic geometrical notions used 
afterwards to prove that in the case of the electroweak interaction the 
moduli space of bosonic vacua is nonempty iff the electromagnetic gauge bundle is
trivial. Moreover, in this case the principal ${\rm SU(2)}\times{\rm U(1)}$ bundle 
underlying the electroweak interaction is also trivial and the moduli space of 
vacua consists of only one point, which is represented by 
$(\Theta_{\!\mbox{\tiny 0}},{\cal V}_{\!\mbox{\tiny 0}})$ iff
the first de Rham cohomology group of space-time is trivial. Thus, if
electromagnetism is supposed to be dynamically generated by spontaneous
symmetry breaking, the corresponding principal ${\rm U(1)}$ bundle
representing the electromagnetic interaction must be trivial. Besides the
usual assumptions of being paracompact, Hausdorff, orientable and smooth 
this statement turns out to be independent of the topology of space-time and 
independent of its geometry.\\

\vspace{0.2cm}

\noindent
{\bf Terminology:} In the following we would like to briefly comment on 
why it might be useful to use a global geometrical description of what is 
usually referred to as an ``elementary particle''. Also, these remarks 
serve to clarify the physical terminology used in this paper.\\
 
In classical physics ``particles'' are geometrically represented by timelike 
(future oriented) one-dimensional submanifolds of a given space-time $\mm$. 
In contrast, in the semi-classical approximation of a (quantum) field theoretical 
description of a ``particle'' the latter is usually identified by its state, described 
by a (quantum) field. Such an identification seems inappropriate since, for
example, the same particle may approach different states. In particular, within 
the realm of gauge theories the state of a particle is supposed to be a gauge 
dependent concept and thus is of no direct physical meaning. Moreover, in a 
quantum field theoretical description of a ``particle'' the notion of the latter
becomes even more subtle for particles may be ``created'', ``annihilated'' or
``transformed into each other''. Consequently, physical notions like ``mass'' 
or ``charge'' usually refer to ``asymptotically free particles''. However, how
one can make the latter geometrically precise within the context of gauge
theories for ``freeness'' means no interaction and thus seems to be a gauge
dependent concept? Also, in particle physics certain asymptotically free particles 
are considered to perform a ``particle multiplet'' which transforms according to 
some (unitary) representation of a given ``gauge group'' G. Again, when seen from 
a gauge geometrical viewpoint such an interpretation of the ``internal space'' 
always refers to a (local) trivialization of the gauge bundle with structure group G.
However, since a (local) trivialization of a gauge bundle ${\cal P}$ cannot be 
performed experimentally such a description of asymptotically free particles within 
gauge theories seems spurious. Notice that this is quite in contrast to relativity, 
where the mathematical concept of a local trivialization has a direct physical 
meaning and so the typical fiber of the tangent bundle of space-time, too. As a
consequence, the concept of an asymptotically free particle should be a purely 
geometrical one. It seems natural to geometrically describe elementary particles, 
at least in a ``semi-classical approximation'' of a quantum theory, as (isomorphism 
classes of) Hermitian vector bundles $\xi$ (see, for instance, \cite{derd'92}). The 
possible states of a particle may then be represented by sections of the appropriate 
bundles. In contrast to a ``particle'' the state of the latter can still be considered as a 
local concept. The gauge interaction between various particles is modeled by the 
assumption that the vector bundles are associated with a given principal G-bundle 
${\cal P}$.\\

For instance, in the case of the bosonic sector of the electroweak interaction the 
``particle content'' of the latter is known not to be given by $(W, B,\Phi)$ but
instead by the ``electromagnetic gauge boson'' $A_{\mbox{\tiny elm}}$ 
together with the massive and electrically (un-)charged
``weak vector bosons'' $Z^0,\,W^\pm$ and the ``physical Higgs boson''
$\Phi_{\mbox{\tiny H,phys}}$. Here,
\bb
\label{weindeco}
A_{\mbox{\tiny elm}} &:=& \cos\!\theta_{\mbox{\tiny W}}\, B
+ \sin\!\theta_{\mbox{\tiny W}}\, W_3,\cr
Z^0&:=& \cos\!\theta_{\mbox{\tiny W}}\, W_3
- \sin\!\theta_{\mbox{\tiny W}}\, B,\cr
W^\pm &:=& W_1 \pm iW_2
\ee
with $W=(W_1,W_2,W_3)\in\Omega^1(\mm,\rr^{\mbox{\tiny 3}})$ the 
``weak gauge boson'' and $\Phi_{\mbox{\tiny H,phys}}$ the ``physical component'' 
of the ``Higgs boson'' $\Phi=(\Phi_{\mbox{\tiny G}},\Phi_{\mbox{\tiny H,phys}})\in
\Omega^0(\mm,\rr^{\mbox{\tiny 4}})$. In the semi-classical approximation a
common usage of terminology in particle physics is that physically non-interacting 
particles are identified with ``free fields'' on space-time $\mm$ (see standard texts, 
for example, \cite{Ait et al'82}, \cite{Nac'89}, or Chapter 21.3 in \cite{Wei'01}). 
However, neither the definition (\ref{weindeco}), nor the notion of a ``free field'', 
in general cannot be gauge invariantly defined. One may thus ask for the 
geometrical meaning of the ``particle content''
\bb
\label{physparticle}
(A_{\mbox{\tiny elm}}, Z^0, W^\pm,\Phi_{\mbox{\tiny H,phys}})
\ee
of the (bosonic part of the) electroweak sector of the Standard Model.\\ 

As it turns out, the ``free particles'' (\ref{physparticle}) are intimately related 
to the notion of a bosonic vacuum of the electroweak interaction. Also, the 
``free particles'' actually have a simple geometrical meaning. Indeed, we shall show 
how the particle content (\ref{physparticle}) can geometrically be considered as 
real line bundles over space-time which naturally come with the geometry of 
spontaneous symmetry breaking of the electroweak interaction. Moreover, these 
line bundles define the extrinsic curvature of the vacuum geometrically considered 
as specific submanifolds. As one may expect, these extrinsic curvatures are 
proportional to the masses of the bosons.\\

However, given such a global description of an elementary particle one may ask 
about the topological structure of the bundles $\xi$. Since they are associated 
bundles, this raises the question about the topology of the underlying gauge 
bundle ${\cal P}$ which, of course, is closely linked to the topology of
space-time $\mm$ itself. In elementary particle physics one usually encounters
$\mm\simeq\rr^{\mbox{\tiny 1,3}}$. Of course, this specific assumption
leads to a definite answer concerning the topology of ${\cal P}$ and thus of $\xi$. 
However, as we have mentioned before when trying to quantize
electromagnetism on a general
space-time $\mm$, one has to consider $dF_{\!\mbox{\tiny elm}}=0$ which,
in general, gives rise only to the local existence of an electromagnetic gauge 
potential $A_{\mbox{\tiny elm}}$. On the other hand, if ${\cal P}$ is supposed
to be trivial, then every gauge potential can be considered as a globally defined
object. But what do we know about the topology of the underlying gauge
bundle? Since the latter has no direct physical meaning it seems inappropriate
to make any a priori assumptions with respect to the topology of ${\cal P}$.
Therefore, our ``strategy'' is the following; the topology of ${\cal P}$ is supposed 
to be arbitrary but fixed, analogous to the assumption of an arbitrary but fixed 
space-time background ${\mm}$. Then, we try to use physically well-established 
assumptions in order to restrict the topological structure of both space-time 
and of the gauge bundle. In the present paper the physically well-motivated 
assumptions made, for example, basically consist in the existence 
of a non-trivial ground state of the Higgs boson and in the assumption that the 
${\rm W}^\pm-$vector bosons of the weak interaction are charge conjugate to 
each other. As we shall see these two physical assumptions fully fix the topology 
of the bundles under consideration.

\section{The geometrical setup}
In this section we summarize the basic geometrical notions which are used to 
generalize $(\Theta_{\!\mbox{\tiny 0}},{\cal V}_{\!\mbox{\tiny 0}})$ to the case 
of arbitrary principal G-bundles ${\cal P}$
\bb
\pi_{\mbox{\tiny P}}:\,{\rm P}&\longrightarrow&\mm\cr
p&\mapsto&x.
\ee
Here, G denotes a finite dimensional compact, semi-simple real Lie group and
$(\mm,g_{\mbox{\tiny M}})$ a smooth semi-Riemannian manifold of arbitrary
signature. Topologically, $\mm$ is supposed to be paracompact, Hausdorff and
orientable. Notice that, like the (semi-)Rie\-mann\-ian structure 
$g_{\mbox{\tiny M}}$, the bundle structure of ${\cal P}$ is supposed to be given 
but otherwise arbitrary.\\

Then, a YMH-gauge theory can be characterized by the following data
\bb
\label{ymhdata}
({\cal P},\rho_{\mbox{\tiny H}},{\rm V}_{\!\mbox{\tiny H}}),
\ee 
where $\rho_{\mbox{\tiny H}}:\,{\rm G}\rightarrow{\rm GL}({\rm N},\cc)$ is a 
unitary representation and 
${\rm V}_{\!\mbox{\tiny H}}:\,\cc^{\mbox{\tiny N}}\rightarrow\rr$ denotes a
G-invariant smooth function that is bounded from below. Moreover, its Hessian is 
supposed to be positive definite transversally to the orbit of each minimum of 
${\rm V}_{\!\mbox{\tiny H}}$. Accordingly, we call such a function 
${\rm V}_{\!\mbox{\tiny H}}$ a ``generalized Higgs potential''.\\

Naturally associated with the data $(\ref{ymhdata})$ are two Hermitian vector
bundles; the ``Higgs bundle'' $\xi_{\mbox{\tiny H}}$ and the ``Yang-Mills bundle''
$\xi_{\mbox{\tiny YM}}\equiv\tau^*_{\mbox{\tiny M}}\ot{\goth{ad}}({\cal P}).$ 
Here, the Higgs bundle is defined by
\bb
\pi_{\mbox{\tiny H}}:\,{\rm E}_{\mbox{\tiny H}}:=
{\rm P}\times_{\rho_{\rm H}}\cc^{\rm N}&\longrightarrow&\mm\cr
{\goth z}\equiv[(p,{\bf z})]&\mapsto&\pi_{\mbox{\tiny P}}(p)
\ee
and the YM bundle as the tensor product of the cotangent bundle 
$\tau^*_{\mbox{\tiny M}}$ of $\mm$ with the ``adjoint bundle'' 
${\goth{ad}}({\cal P})$
\bb
\pi_{\mbox{\tiny ad}}:\,{\rm ad}({\rm P}):=
{\rm P}\times_{\rm ad}{\rm Lie(G)}&\longrightarrow&\mm\cr
\tau\equiv[(p,{\rm T})]&\mapsto&\pi_{\mbox{\tiny P}}(p).
\ee
The Higgs bundle and the Yang-Mills bundle are regarded to geometrically
represent, respectively, the Higgs boson and the Yang-Mills boson. Accordingly,
one may physically interpret the sections of these bundles as the states of the 
respective bosons.\\

Each minimum ${\bf z}_0\in\cc^{\mbox{\tiny N}}$ of the Higgs potential gives
rise to a specific fiber subbundle $\xi_{\mbox{\tiny orb}}$ of the Higgs bundle
called the ``Orbit bundle'' with respect to the minimum ${\bf z}_0$. It is defined
by
\bb
\pi_{\mbox{\tiny orb}}:\,{\cal O}rbit({\bf z}_0):=
{\rm P}\times_{\rho_{\rm orb}}{\rm orbit}({\bf z}_0)&\longrightarrow&\mm\cr
{\goth z}\equiv[(p,{\bf z})]&\mapsto&\pi_{\mbox{\tiny P}}(p),
\ee
where $\rho_{\mbox{\tiny orb}}:=
\rho_{\mbox{\tiny H}}|_{\mbox{\tiny${\rm orbit}({\bf z}_0)$}}$. Notice that
$\xi_{\mbox{\tiny orb}}\simeq\xi'_{\mbox{\tiny orb}}$ iff ${\bf z}'_0$ and
${\bf z}_0$ are on the same orbit. Like the gauge bundle ${\cal P}$, the orbit 
bundle has no direct physical meaning. However, since 
$\xi_{\mbox{\tiny orb}}\subset\xi_{\mbox{\tiny H}},$ any section ${\cal V}$ of
the orbit bundle physically represents a possible ground state of the Higgs
boson. We therefore call ${\cal V}$ a ``vacuum section''. We denote by
${\cal H}\subset{\cal G}$ the invariance group of the vacuum section. It is
a closed subgroup of the gauge group ${\cal G}$ of ${\cal P}$ and may point-wise 
be identified with the isotropy group ${\rm I}({\bf z}_0)\subset{\rm G}$
of the minimum ${\bf z}_0$.\\

Every vacuum section singles out a specific class of connections on ${\cal P}$.
For this we remark that each vacuum section ${\cal V}$ is in one-to-one
correspondence to an ``H-reduction'' $({\cal Q},\iota)$ of the principal
G-bundle ${\cal P}$ (see, e. g., \cite{koba/nomi}). That is, there is a unique 
principal H-bundle ${\cal Q}$
\bb
\pi_{\mbox{\tiny Q}}:\,{\rm Q}&\longrightarrow&\mm\cr
q&\mapsto&x
\ee
together with a bundle embedding $\iota:\,{\cal Q}\hookrightarrow{\cal P}$
(i. e. $\pi_{\mbox{\tiny P}}(\iota(q))=\pi_{\mbox{\tiny Q}}(q)$ for all $q\in{\rm Q}$),
such that ${\rm H}\simeq{\rm I}({\bf z}_0)$. Indeed, in contrast to the more
physically intuitive notion of a vacuum section the usually geometrical description
of spontaneous symmetry breaking only refers to the notion of a bundle
reduction (see, e. g., \cite{bleecker}, \cite{choquet eta}, \cite{sternberg}, 
\cite{trautmann}.)\\
 
Note that, in general, the principal H-bundle ${\cal Q}$ will be non-trivial even
if the principal G-bundle ${\cal P}$ is equivalent to the trivial one. Of course,
the triviality of ${\cal Q}$ implies the triviality of ${\cal P}$. Also, any connection 
on ${\cal Q}$ generally induces a connection on ${\cal P}$ but not vice versa.
A connection ${\cal A}$ on ${\cal P}$ is said to be ``H-reducible'' iff 
$\iota^*{\cal A}$ is also a connection on ${\cal Q}$. In this case we call ${\cal A}$
compatible with the appropriate vacuum section ${\cal V}$. A simple
criterion for a connection to be compatible with a vacuum section is given by
the following\\

\begin{lemma}
A connection on ${\cal P}$ is compatible with a vacuum section
${\cal V}\in\Gamma(\xi_{\mbox{\tiny orb}})$ iff the associated connection 
${\cal A}\in{\cal A}(\xi_{\mbox{\tiny H}})$ on the Higgs bundle satisfies
\bb
\label{hred}
d_{\!\mbox{\tiny A}}{\cal V}=0
\ee
with $d_{\!\mbox{\tiny A}}$ the exterior covariant derivative with respect 
to ${\cal A}$.
\end{lemma}

\vspace{0.5cm}

\noindent
{\bf Proof:} Since a connection on ${\cal P}$ is H-reducible iff the corresponding 
connection form on ${\cal P}$ takes values in Lie(H) the statement
follows from ${\cal V}(x) =
[(\iota(q),{\bf z}_0)]|_{\mbox{\tiny$q\in\pi_{\rm Q}^{-1}(x)$}}$.\hfill$\Box$\\

\vspace{0.5cm}

We call a YMH pair $({\Theta},{\cal V})\in
{\cal A}(\xi_{\mbox{\tiny H}})\times\Gamma(\xi_{\mbox{\tiny H}})$ a 
(classical) ``bosonic vacuum'' (or, in the context of this paper, an ``electroweak 
vacuum'') iff ${\cal V}$ denotes a vacuum section with respect to some chosen 
minimum ${\bf z}_0$ and $\Theta$ a connection on $\xi_{\mbox{\tiny H}}$ 
associated with a flat connection on ${\cal P}$ and which is compatible 
with ${\cal V}$.\\

The notion of a (classical) bosonic vacuum introduced here indeed generalizes 
the geometrical interpretation of a vacuum as described in the introduction.
Since in the case of $(\mm,g_{\mbox{\tiny M}})\simeq\rr^{\mbox{\tiny 1,3}}$  
it follows that both ${\cal P}$ and ${\cal Q}$ must be trivial for any vacuum
section. The latter may then be identified with smooth mappings
$\nu:\,\mm\rightarrow{\rm orbit}({\bf z}_0)$. Moreover, any such mapping
is easily shown to be gauge equivalent to the canonical mapping 
${\cal V}_{\!\mbox{\tiny 0}}$ which corresponds to the canonical embedding
\bb
\mm\times{\rm H}&\hookrightarrow&\mm\times{\rm G}\cr
(x,h)&\mapsto&(x,h).
\ee

In \cite{tolk1'03} it is shown that one encounters a similar situation in
the case where $\pi_1(\mm)=0$. More precisely, in the given reference it is
proved that on a simply connected manifold $\mm$ there exists at most one
vacuum for each orbit. Moreover, these vacua are all gauge equivalent to 
$(\Theta_{\!\mbox{\tiny 0}},{\cal V}_{\!\mbox{\tiny 0}})$. It is then a natural 
question to ask for the structure of the 
moduli space of vacua in the case of $\pi_1(\mm)\not=0$. This will be done in 
the next section for the particular case of the electroweak interaction.\\

We close this section with the remark that with respect to any vacuum
section ${\cal V}$ the (realification of the) Higgs bundle decomposes into
the Whitney sum of two real subbundles called the ``Goldstone bundle''
$\xi_{\mbox{\tiny G}}$ and the ``physical Higgs bundle'' 
$\xi_{\mbox{\tiny H,phys}}$, i. e.
\bb
\xi_{\mbox{\tiny H}} = \xi_{\mbox{\tiny G}}\op\xi_{\mbox{\tiny H,phys}}.
\ee

Moreover, since a vacuum section can be considered as a specific embedding of 
space-time into the total space of the Higgs bundle, the tangent bundle of $\mm$ 
together with the Goldstone and the physical Higgs bundle build a ``global 3-Bein'' 
along the vacuum ${\cal V}(\mm)\subset{\rm E}_{\mbox{\tiny H}}$. In particular, 
the physical Higgs bundle $\xi_{\mbox{\tiny H,phys}}$ can be identified with the 
normal bundle of ${\cal O}rbit({\bf z}_0)\subset{\rm E}_{\mbox{\tiny H}}$ restricted 
to ${\cal V}(\mm)\subset{\rm E}_{\mbox{\tiny H}}$. Notice that the 3-Bein is
orthogonal with respect to the metric $g_{\mbox{\tiny H}}$ induced
by $(\Theta, g_{\mbox{\tiny M}})$ on ${\rm E}_{\mbox{\tiny H}}$. The exterior
curvature of ${\cal O}rbit({\bf z}_0)$ along the vacuum is proportional to the
mass of the physical Higgs boson.

\section[The moduli space of vacua]{The moduli space of vacua of the electroweak 
interaction}
The structure of the moduli space of the bosonic vacua is found to be surprisingly 
simple in the case of the electroweak interaction. This is so because this interaction 
turns out to have some special topological features which we will discuss in this 
section.\\

First, we again summarize the data defining the electroweak interaction as a 
specific YMH gauge theory. In this case 
$({\cal P},\rho_{\mbox{\tiny H}},{\rm V}_{\!\mbox{\tiny H}})$ is given by
\begin{itemize}
\item a principal ${\rm G}:={\rm SU(2)}\times{\rm U(1)}-$bundle ${\cal P}$, 
\item the unitary representation 
$\rho_{\mbox{\tiny H}}:\,{\rm G}\rightarrow{\rm GL}(2,\cc),\;
g\equiv(g_{\mbox{\tiny(2)}},g_{\mbox{\tiny(1)}})\mapsto
g_{\mbox{\tiny(2)}}\,g_{\mbox{\tiny(1)}}^{\mbox{\tiny y}}$ (with 
``hypercharge'' ${\rm y}\in\qq$),
\item the Higgs potential 
${\rm V}_{\!\mbox{\tiny H}}(\bf z):=\lambda\,|{\bf z}|^{\mbox{\tiny 4}}-
\mu^{\mbox{\tiny 2}}\,|{\bf z}|^{\mbox{\tiny 2}}\;(\lambda,\mu>0)$.
\end{itemize}

As already mentioned in the introduction, this Higgs potential has but
one orbit of minima that is isomorphic to 
${\rm S}^{\mbox{\tiny 3}}\subset\cc^{\mbox{\tiny 2}}$. Moreover, it has the
special feature of being ``rotationally symmetric''. That is, the Higgs potential 
can be written as ${\rm V}_{\!\mbox{\tiny H}} = f_{\mbox{\tiny H}}\circ r,$ with
$r({\bf z}):=|{\bf z}|$ the radial function and 
$f_{\mbox{\tiny H}}\in{\cal C}^\infty(\rr_+)$ bounded from below. As a
consequence, it can be shown that there exists a vacuum section ${\cal V}$
iff the Higgs bundle $\xi_{\mbox{\tiny H}}$ admits a non-vanishing section.
Moreover, with respect to such a vacuum section the physical Higgs bundle
is a trivial real line bundle. Also, for any non-vanishing section $\Phi$ of the 
Higgs bundle one may always find a vacuum, such that $\Phi$ can be identified 
with a section of the corresponding physical Higgs bundle (see again \cite{tolk1'03}).
In other words, there always exists a vacuum such that $\Phi$ is in the
``unitary gauge'' with respect to this vacuum.\\

Next, we prove that in the case of the electroweak interaction the adjoint bundle 
${\goth{ad}}({\cal P})$ decomposes into the Whitney sum of two real line bundles 
and one real vector bundle of rank two.\\

\begin{proposition}
Let $({\cal P},\rho_{\mbox{\tiny H}},{\rm V}_{\!\mbox{\tiny H}})$ be the data
defining the electroweak interaction as a YMH gauge theory. Also, let
${\cal V}\in\Gamma(\xi_{\mbox{\tiny orb}})$ be a vacuum section with respect
to some minimum ${\bf z}_0$ of the Higgs potential. With respect to the 
vacuum section ${\cal V}$ the adjoint bundle ${\goth{ad}}(\cal P)$, considered
as a vector bundle, decomposes as
\bb
\label{gwshiggsdinner}
{\goth{ad}}({\cal P})\simeq
{\goth{ad}}({\cal Q})\op(\xi_{\mbox{\tiny Z}}\op\xi_{\mbox{\tiny W}}).
\ee
Here, $\xi_{\mbox{\tiny Z}}$ and $\xi_{\mbox{\tiny W}}$ respectively denote a
real vector bundle of rank one and of rank two.
\end{proposition}

\vspace{0.2cm}

\noindent
{\bf Proof:} When considered as a vector bundle ${\goth{ad}}(\cal P)$ is 
H-reducible and decomposes as (see, \cite{tolk1'03})
\bb
\label{higgsdinner}
{\goth{ad}({\cal P})}\simeq{\goth{ad}({\cal Q})}\op\xi_{\mbox{\tiny G}}.
\ee
Likewise, the Yang-Mills mass matrix
\bb
\label{ymm}
{\cal V}^*{\rm M}^2_{\!\mbox{\tiny YM}}:\,
{\goth{ad}({\cal P})}&\longrightarrow&{\goth{ad}({\cal P})}\cr
\tau\equiv[(p,{\rm T})]&\mapsto&{\cal V}^*{\rm M}^2_{\!\mbox{\tiny YM}}(\tau),
\ee
with
${\cal V}^*{\rm M}^2_{\!\mbox{\tiny YM}}(\tau)(x):=
[(p,{\rm ad}_{g^{-1}}({\rm\bf M}^2_{\!\mbox{\tiny YM}}({\bf z}_0)
{\rm ad}_g({\rm T})))]|_{\mbox{\tiny$p\in\pi_{\rm P}^{-1}(x)$}},$ decomposes as
\bb
{\cal V}^*{\rm M}^2_{\!\mbox{\tiny YM}} =
(0)\op{\cal V}^*{\rm M}^2_{\!\mbox{\tiny YMG}},
\ee
where ${\cal V}^*{\rm M}^2_{\!\mbox{\tiny YMG}}:=
{\cal V}^*{\rm M}^2_{\!\mbox{\tiny YM}}|_{\mbox{\tiny$\xi_{\rm G}$}}$ has
maximal rank. In (\ref{ymm}) $p=\iota(q)g$ for arbitrary 
$q\in\pi_{\mbox{\tiny Q}}^{-1}(x)$, $g\in{\rm G}$ and 
${\rm\bf M}^2_{\!\mbox{\tiny YM}}({\bf z}_0)\in{\rm End}({\rm Lie(G)})$ is defined 
by $\beta({\rm\bf M}^2_{\!\mbox{\tiny YM}}({\bf z}_0)({\rm T}),{\rm T}')=
2\,\rho'_{\mbox{\tiny H}}({\rm T}){\bf z}_0\cdot
\rho'_{\mbox{\tiny H}}({\rm T}'){\bf z}_0$ for all
${\rm T},{\rm T}'\in{\rm Lie(G)}$. The symmetric bilinear form $\beta$
on Lie(G) is given by the most general parameterized Killing form, and 
$\rho'_{\mbox{\tiny H}}$ denotes the real form of the ``derived'' representation
(see again, loc sit). Note that ${\cal V}^*{\rm M}^2_{\!\mbox{\tiny YM}}$ has
constant spectrum and lays within the commutant of the reduced gauge group 
${\cal H}$. Moreover, the spectrum only depends on the orbit of ${\bf z}_0$ and 
not on the vacuum section ${\cal V}$ chosen. 
From the above follows that one can decompose the
Goldstone bundle $\xi_{\mbox{\tiny G}}$ into the eigenbundles of the Yang-Mills
mass matrix. Since the latter commutes with the (representation of the)
electromagnetic gauge group, the spectrum of
${\cal V}^*{\rm M}^2_{\!\mbox{\tiny YM}}$ consists of maximally two different 
eigenvalues ${\rm m}_{\mbox{\tiny Z}},{\rm m}_{\mbox{\tiny W}}\in\rr_+$.
If we denote by $\rho_{\mbox{\tiny G}}$ the restriction of the real form of
$\rho_{\mbox{\tiny H}}$ to the typical fiber of the Goldstone bundle, then
$\rho_{\mbox{\tiny G}}(h)={\bf A}\op(1)$ for all $h\in{\rm H}$
(where ${\bf A}\in{\rm SO(2)}$). Therefore, the Goldstone bundle decomposes 
into the Whitney sum of a real rank one vector bundle $\xi_{\mbox{\tiny Z}},$ 
which corresponds to the eigenvalue ${\rm m}_{\mbox{\tiny Z}},$ and a
rank two vector bundle $\xi_{\mbox{\tiny W}},$ which corresponds to the
eigenvalue ${\rm m}_{\mbox{\tiny W}}$ of the Yang-Mills mass matrix.\hfill$\Box$\\

The ${\cal V}-$induced isomorphism (\ref{higgsdinner}) can be considered as
a geometrical variant of what is called the ``Higgs Dinner'' (see, \cite{higgs'64}).
As a consequence, in the case of the electroweak interaction the Yang-Mills 
bundle decomposes as
\bb
\xi_{\mbox{\tiny YM}} \simeq \xi_{\mbox{\tiny elm}}\op\,
(\xi_{\mbox{\tiny${\rm W}^\pm$}}\op\,\xi_{\mbox{\tiny${\rm Z}^0$}}),
\ee
where $\xi_{\mbox{\tiny elm}}:=
\tau^*_{\mbox{\tiny M}}\ot{\goth{ad}({\cal Q})}$ geometrically represents
the electromagnetic gauge boson and $\xi_{\mbox{\tiny${\rm Z}^0$}}:=
\tau^*_{\mbox{\tiny M}}\ot\xi_{\mbox{\tiny Z}}$ the massive electromagnetically
neutral ${\rm Z}^0$ vector boson of the weak interaction. Note that 
${\goth{ad}}({\cal Q})$ and $\xi_{\mbox{\tiny${\rm Z}$}}$ are trivial, for they are 
``uncharged'' (i. e. they carry the trivial representation of the electromagnetic 
gauge group ${\cal H}$).\\

The rank two  vector bundle 
\bb
\label{wboson}
\xi_{\mbox{\tiny${\rm W}^\pm$}}:=\tau^*_{\mbox{\tiny M}}\ot\xi_{\mbox{\tiny W}}
\ee
geometrically represents a massive electrically charged vector boson. Indeed,
one may naturally identify the real vector bundle $\xi_{\mbox{\tiny W}}$ of
rank two with the complex line bundle associated with the electromagnetic gauge
bundle ${\cal Q}$ via the fundamental representation of ${\rm H}$. Physically one 
can thus identify (\ref{wboson}) either with the $W^+-$ or with the $W^--$boson 
of the weak interaction. Because ${\bf A}\in{\rm SO(2)}$ has no real 
eigenvalues the real vector bundle $\xi_{\mbox{\tiny${\rm W}$}}$ does not 
naturally decompose into two real line bundles that geometrically represent 
the $W^+-$ and the $W^--$boson. The motivation for nonetheless identifying 
$\xi_{\mbox{\tiny${\rm W}^\pm$}}$ with either of the well-known massive 
electrically charged vector bosons is as follows: Considered as a complex
vector bundle of rank two $\cc\ot\xi_{\mbox{\tiny W}}$ decomposes into
\bb
\label{compl}
\cc\ot\xi_{\mbox{\tiny W}}=
\xi_{\mbox{\tiny${\rm W}^+$}}\op\,\xi_{\mbox{\tiny${\rm W}^-$}}.
\ee
Here, either $\xi_{\mbox{\tiny${\rm W}^+$}}$ is assumed to carry the 
fundamental representation of the electromagnetic gauge group and then 
$\xi_{\mbox{\tiny${\rm W}^-$}}={\overline{\xi_{\mbox{\tiny${\rm W}^+$}}}}$ or
vice versa. Both complex line bundles are also eigenbundles of the Yang-Mills 
mass matrix with respect to the eigenvalue ${\rm m}_{\mbox{\tiny W}}$. Since 
$\xi_{\mbox{\tiny W}}$ carries the fundamental representation of the 
(real form of the) electromagnetic gauge group one may naturally identify 
$\xi_{\mbox{\tiny W}}$ either with
$\xi_{\mbox{\tiny${\rm W}^+$}}$ or with $\xi_{\mbox{\tiny${\rm W}^-$}}$.\\

As mentioned before, to geometrically represent both the $W^+-$ and the
$W^--$boson as subbundles of the Yang-Mills bundle one needs additional 
structure. Physically, this additional piece of input arises from the assumption 
that the $W^\pm-$bosons are charge conjugate to each other. Since in the
case at hand charge conjugation is the same as complex conjugation on the
complex line bundle $\xi_{\mbox{\tiny W}},$ charge conjugation geometrically 
means that there exists a real line bundle such that its complexification equals 
$\xi_{\mbox{\tiny W}}$. In other words, to assume that the $W^\pm-$bosons of
the weak interaction are charge conjugate to each other is the same as
to assume that there exist real line bundles 
$\xi_{\mbox{\tiny${\rm W}_1$}}\simeq\xi_{\mbox{\tiny${\rm W}_2$}}$, such that
\bb
\xi_{\mbox{\tiny W}} = 
\xi_{\mbox{\tiny${\rm W}_1$}}\op\,\xi_{\mbox{\tiny${\rm W}_2$}}.
\ee
In this case, the Yang-Mills mass matrix
${\cal V}^*{\rm M}^2_{\!\mbox{\tiny YM}}$ together with the real structure ${\cal J}$
(complex conjugation) on $\xi_{\mbox{\tiny W}}$ permits to decompose 
the Yang-Mills bundle into the Whitney sum of four real line bundles
\bb
\xi_{\mbox{\tiny YM}} = \xi_{\mbox{\tiny elm}}\op\,
(\xi_{\mbox{\tiny${\rm Z}^0$}}\op\,
\xi_{\mbox{\tiny${\rm W}_1$}}\op\,\xi_{\mbox{\tiny${\rm W}_2$}}),
\ee
where
\bb
\label{cwboson}
\xi_{\mbox{\tiny${\rm W}^+$}} &=&
\xi_{\mbox{\tiny${\rm W}_1$}}\op\, i\xi_{\mbox{\tiny${\rm W}_2$}},\cr
\xi_{\mbox{\tiny${\rm W}^-$}} &=&
\xi_{\mbox{\tiny${\rm W}_1$}}\ominus\, i\xi_{\mbox{\tiny${\rm W}_2$}}.
\ee

Note that it is a well-established empirical fact that the electromagnetic 
interaction is invariant with respect to charge conjugation\footnote{That is,
the electromagnetic interaction does not permit to absolutely distinguish 
between particles and anti-particles.}. Therefore,
to assume the existence of charge conjugation is physically well-motivated. 
The point here is that charge conjugation comes within the bosonic sector of the 
Standard Model since spontaneous symmetry breaking not only creates massive 
but also charged bosons\footnote{By ``charge'' we always mean 
``electromagnetic charge''. We carefully distinguish between the notions of
``charge'' and ``gauge coupling constant''. The former is a dynamically conserved
quantity due to Noether's theorem, whereas the latter is conserved by construction 
(it simply parameterizes the most general Killing form on Lie(G)).}.\\

Of course, the decomposition (\ref{cwboson}) is the geometrical analogy to 
the usual complex linear combination of the electroweak bosons found in the
literature on the Standard Model (see, eq. \ref{weindeco}). Like in the local 
description the global decomposition (\ref{cwboson}) is unique, i. e. the 
correspondence between $(\xi_{\mbox{\tiny$W^+$}},\xi_{\mbox{\tiny$W^+$}})$ 
and $(\xi_{\mbox{\tiny${\rm W}_1$}},\xi_{\mbox{\tiny${\rm W}_2$}}$) is
one-to-one. Indeed, we have the following\\

\begin{lemma}
On the complex line bundle $\xi_{\mbox{\tiny W}}$ there exists a complex
conjugation iff it is trivial.
\end{lemma}

\vspace{0.2cm}

\noindent
{\bf Proof:} The statement follows from the fact that a complex vector bundle $\xi$
of rank N possesses a complex conjugation (i. e. a real structure) iff all of its odd 
Chern classes 
$c_{\mbox{\tiny 2k+1}}(\xi)\in{\rm H}^{4k+2}_{\mbox{\tiny deR}}(\mm)$ vanish.
Though this is hard to prove in general, for N=1 the proof is elementary. Indeed,
let $\xi_{\mbox{\tiny W}}$ be trivial. Then, the structure group can be reduced
to the identity and $\xi_{\mbox{\tiny W}}$ possesses a canonical complex conjugation.
If we let the complex line bundle $\xi_{\mbox{\tiny W}}$ be equipped with a
complex conjugation,  $\xi_{\mbox{\tiny W}}$ is the complexification of a
real line bundle. When considered as a real vector bundle $\xi_{\mbox{\tiny W}}$
decomposes into the Whitney sum of two real line bundles. However,
since all one-dimensional representations of SO(2) are trivial it follows that each 
of the real line bundles is trivial.\hfill$\Box$\\ 

\vspace{0.2cm}

As a consequence we conclude that with respect to a bosonic vacuum
$(\Theta,{\cal V})$ the Yang-Mills bundle of the electroweak interaction 
reads\footnote{ Because of the tensor product with the cotangent bundle, 
the Yang-Mills bundle $\xi_{\mbox{\tiny YM}}$ is always considered 
as a real vector bundle. Thus, to consider the Yang-Mills bundle as the 
tensor product of $\tau^*_{\mbox{\tiny M}}$ with (\ref{compl}) would not
make sense.}
\bb
\label{ymg}
\xi_{\mbox{\tiny YM}}\simeq\bigoplus^4\tau^*_{\mbox{\tiny M}}.
\ee
Next, we show that this fully fixes the topological structure of both the 
electromagnetic and the electroweak gauge bundle.\\

\begin{proposition}
The gauge bundles underlying the electroweak interaction and electromagnetism 
are trivial.
\end{proposition}

\vspace{0.2cm}

\noindent
{\bf Proof:} Since $\xi_{\mbox{\tiny W}}$ is a complex line bundle carrying
the fundamental representation of the electromagnetic gauge group 
its frame bundle can be canonically identified with ${\cal Q}$. Since there 
exists a complex conjugation on $\xi_{\mbox{\tiny W}}$ iff $\xi_{\mbox{\tiny W}}$ 
is trivial, the principal ${\rm H}\equiv{\rm U}_{\mbox{\tiny elm}}(1)-$bundle 
${\cal Q}$ must also be trivial\footnote{Here, the structure group of 
electromagnetism is defined by
$${\rm U}_{\mbox{\tiny elm}}(1)\simeq{\rm I}({\bf z}_0)=
\{h\equiv\exp(\theta[{\rm T} + {\rm y}i])\,|\,{\rm T}={\rm T}({\bf z}_0)\in{\rm su(2)}, 
{\rm tr}([{\rm T} + {\rm y}i]^2)=-1,\,\theta\in\rr\}.$$}. Consequently, as an 
extension of the electromagnetic gauge bundle, the principal 
${\rm SU(2)}\times{\rm U(1)}$ bundle ${\cal P}$ must also be trivial 
(see our corresponding remark of the last section).\hfill$\Box$\\

\vspace{0.2cm}

As a consequence, we conclude that the $W^\pm-$vector bosons of the weak
interaction are charge conjugate to each other iff the electroweak gauge bundle
is trivial. In what follows we will slightly change our argument and present two 
alternative proofs of the triviality of the electroweak gauge bundle. The first 
proof refers to the existence of a bosonic vacuum $(\Theta,{\cal V})$. In contrast,
the second proof only refers to the existence of a non-trivial ground state
${\cal V}\in\Gamma(\xi_{\mbox{\tiny orb}})$ of the Higgs boson.\\

\vspace{0.2cm}

\begin{proposition}
\label{prop}
The electroweak interaction admits a bosonic vacuum iff its underlying gauge 
bundle ${\cal P}$ is trivial.
\end{proposition}

\vspace{0.2cm}

\noindent
{\bf Proof:} Of course, if the principal ${\rm SU(2)}\times{\rm U(1)}$ bundle 
${\cal P}$ is assumed to be trivial, then the YMH pair 
$(\Theta_{\!\mbox{\tiny 0}},{\cal V}_{\!\mbox{\tiny 0}})$ will
serve as a vacuum for all minima ${\bf z}_0\in\cc^{\mbox{\tiny 2}}$. Now, let
$(\Theta,{\cal V})$ be a vacuum with respect to the data defining the 
electroweak interaction as a YMH gauge theory. Then, a principal 
${\rm U}_{\mbox{\tiny elm}}(1)-$bundle ${\cal Q}$ together with a bundle 
embedding $\iota:\,{\cal Q}\hookrightarrow{\cal P}$ uniquely corresponds
to ${\cal V}$ such that $\iota^*\Theta$ is also a
flat connection on the reduced bundle ${\cal Q}$. Thus, the first Chern class
$c_{\mbox{\tiny 1}}({\cal Q})\in{\rm H}^2_{\mbox{\tiny deR}}(\mm)$ of 
the electromagnetic gauge bundle must vanish. Since principal U(1)-bundles 
are classified by their first Chern class it follows that
${\cal Q}$ must be trivial. Since the electromagnetic gauge bundle ${\cal Q}$ is 
regarded as a reduction of the electroweak gauge bundle ${\cal P},$ the latter 
must be also trivial.\hfill$\Box$\\

\vspace{0.2cm}

Like in the case of the assumption $\pi_1(\mm)=0$ the above given argument
makes use of the existence of a flat connection $\Theta$ on ${\cal P}$. While
physically motivated, this assumption turns out to be mathematically very
restrictive. However, taking charge conjugation into account, we may draw 
the same conclusion as above by only referring to the existence of a non-trivial
ground state of the Higgs boson.\\

\vspace{0.2cm}

\begin{proposition}
The Higgs boson of the electroweak sector of the Standard Model possesses a
non-trivial ground state iff the electroweak interaction is geometrically modeled
by the trivial principal ${\rm SU(2)}\times{\rm U(1)}$ bundle. 
\end{proposition}

\vspace{0.2cm}

\noindent
{\bf Proof:} Again, if the electroweak gauge bundle ${\cal P}$ is supposed to be
trivial, then every minimum ${\bf z}_0\in\cc^{\mbox{\tiny 2}}$ gives rise to a 
canonical section of the (also trivial) orbit bundle. To prove the converse, let
${\cal V}\in\Gamma(\xi_{\mbox{\tiny orb}})$ be a section of the orbit bundle
with respect to some chosen minimum ${\bf z}_0\in{\rm S}^{\mbox{\tiny 3}}$.
Also, let $({\cal Q},\iota)$ be the corresponding electromagnetic reduction of the 
electroweak gauge bundle ${\cal P}$. Again, the isomorphism class of the 
electromagnetic gauge bundle ${\cal Q}$ is fully determined by its first Chern class 
$c_{\mbox{\tiny 1}}({\cal Q})\in{\rm H}^2_{\mbox{\tiny deR}}(\mm)$
(c.f., for instance, the appendix of \cite{freed et al'XY}). Therefore, 
electromagnetism is invariant with respect to charge conjugation iff 
$c_{\mbox{\tiny 1}}({\cal Q})=0$.\hfill$\Box$\\

\vspace{0.2cm}

Even though it only refers to the existence of a non-trivial ground state of the 
Higgs boson the above result also implies, of course, the existence of a bosonic 
vacuum as, for instance, the canonical one represented by the YMH pair
$(\Theta_{\!\mbox{\tiny 0}},{\cal V}_{\!\mbox{\tiny 0}})$. Our main theorem 
then says that in the case of the electroweak 
interaction this kind of a bosonic vacuum is in fact the only one, provided
that ${\rm H}^1_{\mbox{\tiny deR}}(\mm)=0$. The proof of
this statement makes use of another special feature of the electroweak 
interaction.\\

\vspace{0.2cm}

\begin{satz}
\label{satz}
Let $({\cal P},\rho_{\mbox{\tiny H}},{\rm V}_{\!\mbox{\tiny H}})$ be the data
defining the electroweak interaction as a YMH gauge theory. Then, the
corresponding moduli space of bosonic vacua ${\goth{M}}_{\mbox{\tiny vac}}$ 
is an affine space with vector space ${\rm H}^1_{\mbox{\tiny deR}}(\mm)$.
\end{satz}

\vspace{0.2cm}

\noindent
{\bf Proof:} According to the above propositions we already know that the moduli 
space of bosonic vacua is non-empty iff the electroweak gauge bundle ${\cal P}$
is trivial. As a consequence, every vacuum section ${\cal V}$ can be identified
with a smooth mapping $\nu:\,\mm\rightarrow{\rm orbit}({\bf z}_0)$. In
\cite{tolk1'03} it was shown, however, that in the case of the electroweak
interaction the principal H-bundle 
\bb
{\rm G}&\longrightarrow&{\rm orbit}({\bf z}_0)\cr
g&\mapsto& g{\bf z}_0
\ee
 is
also trivial. Therefore, every smooth mapping $\nu$ possesses a smooth lift
$\gamma:\,\mm\rightarrow{\rm G}$, such that $\nu(x)=\gamma(x){\bf z}_0$.
In other words, every vacuum section is gauge equivalent to the canonical
vacuum section. Moreover, since the affine space of H-reducible connections 
${\cal A}$ on ${\cal P}$ can be canonically identified with $\Omega^1(\mm)$ it 
follows that each flat connection $\Theta$ uniquely corresponds to an element 
of ${\rm H}^1_{\mbox{\tiny deR}}(\mm)$. If the latter is trivial, then $\Theta$ 
is gauge equivalent to the canonical connection $\Theta_{\!\mbox{\tiny 0}}$ 
and the moduli space of bosonic vacua consists of at most one point represented 
by the canonical Yang-Mills-Higgs pair 
$(\Theta_{\!\mbox{\tiny 0}},{\cal V}_{\!\mbox{\tiny 0}})$.\hfill$\Box$\\

We may thus summarize our main result by
\bb
\label{mainres}
\goth{M}_{\mbox{\tiny vac}} \simeq {\rm H}^1_{\mbox{\tiny deR}}(\mm),
\ee 
iff the electroweak gauge bundle ${\cal P}$ is trivial. This in turn is in one-to-one
correspondence with the assumption of the existence of charge conjugation 
${\cal J}.$ Locally, the relation (\ref{weindeco}) between the ``interacting fields'' 
$(W, B, \Phi)$ and the ``(asymptotically) free fields'' 
$(A_{\mbox{\tiny elm}}, Z^0, W^\pm,\Phi_{\mbox{\tiny H,phys}})$ is unambiguous. 
However, whether this holds also true when seen from a global perspective 
depends on the structure of the moduli space ${\goth{M}_{\mbox{\tiny vac}}}$
of electroweak vacua.\\

Let $\Delta:\mm\rightarrow\mm\times\mm$ be the diagonal embedding 
$x\mapsto(x,x).$ Since the structure group 
${\rm G}\equiv{\rm SU(2)}\times{\rm U(1)}$ of the electroweak gauge bundle
${\cal P}$ is a direct product one obtains\footnote{I would like to thank E. Binz for a
corresponding hint.}
\bb
{\cal P} = \Delta^*({\cal P}_2\times{\cal P}_1).
\ee 
Here, respectively, ${\cal P}_1$ and ${\cal P}_2$ are appropriate principal U(1) 
and principal SU(2) bundles over $\mm$, and $\Delta^*({\cal P}_2\times{\cal P}_1)$
means the pull-back bundle of ${\cal P}_2\times{\cal P}_1$ with respect to $\Delta$. 
According to the Higgs dinner (\ref{higgsdinner}) with respect to an electroweak 
vacuum $(\Theta,{\cal V})\in
{\cal A}(\xi_{\mbox{\tiny H}})\times\Gamma(\xi_{\mbox{\tiny H}})$, we obtain the 
following two orthogonal decompositions of the adjoint bundle ${\goth{ad}({\cal P})}$ 
(and thus of the Yang-Mills bundle $\xi_{\mbox{\tiny YM}}$)
\bb
{\goth{ad}({\cal P})} &=& {\goth{ad}({\cal P}_1)}\op{\goth{ad}({\cal P}_2)}\cr
&\simeq&
{\goth{ad}({\cal Q})}\op\xi_{\mbox{\tiny Z}}\op\xi_{\mbox{\tiny W}}.
\ee
Because of $\xi_{\mbox{\tiny W}}\subset{\goth{ad}({\cal P}_2)},$ there is a unique 
real line bundle $\xi_{\mbox{\tiny${\rm W}_3$}}\subset{\goth{ad}({\cal P}_2)},$ 
such that
\bb
{\goth{ad}({\cal P}_2)}\simeq\xi_{\mbox{\tiny W}}\op\xi_{\mbox{\tiny${\rm W}_3$}}.
\ee

The ``electroweak mixing angle'' $\theta_{\mbox{\tiny W}}$ of (\ref{weindeco}) is 
geometrically represented by the isometric isomorphism (over the identity on 
$\mm$)
\bb
\label{weakangle}
{\goth{ad}({\cal P}_1)}\op\xi_{\mbox{\tiny${\rm W}_3$}}\simeq
{\goth{ad}({\cal Q})}\op\xi_{\mbox{\tiny Z}}.
\ee

Note that both sides of (\ref{weakangle}) are orthogonal complements of 
$\xi_{\mbox{\tiny W}}\subset{\goth{ad}({\cal P})}.$ Also note that the isomorphism
(\ref{weakangle}) does not take into account the triviality of the electroweak gauge
bundle. The triviality of ${\cal P}$ only corresponds to the last relation of 
(\ref{weindeco}) which is geometrically described by (\ref{cwboson}). Of course,
the isomorphism (\ref{weakangle}) only depends on 
$[({\Theta},{\cal V})]\in{\goth{M}}_{\mbox{\tiny vac}}$.\\

If space-time is assumed to be simply connected, then, up to gauge equivalence, 
there is at most one non-trivial bosonic vacuum. This, of course, holds 
true for $(\mm,g_{\mbox{\tiny M}})\simeq\rr^{\mbox{\tiny 1,3}}$,
usually encountered in particle physics. This also fits in with the corresponding 
results presented in \cite{tolk1'03} for general YMH data
$({\cal P},\rho_{\mbox{\tiny H}},{\rm V}_{\!\mbox{\tiny H}}),$ where, we 
showed that $\pi_1(\mm)=0$ implies that the moduli space of bosonic vacua
consists of at most one point. In the case considered in the present paper, 
however, the topology of space-time is two-fold related to the bosonic vacua; 
the existence is tied to ${\rm H}^2_{\mbox{\tiny deR}}(\mm)$, whereas 
uniqueness is related to ${\rm H}^1_{\mbox{\tiny deR}}(\mm)$. The
non-triviality of the latter may physically be interpreted in terms of the
Ahoronov-Bohm effect. The anholonomy of the electron's phase uncovers
the non-triviality of the electroweak vacuum.\\

\section[Asymptotically Free Particles]{Asymptotically Free Particles}
In this section we summarize the geometrical meaning of the physical notion of
 a ``free particle'' in the realm of gauge theory. The motivation for this is as 
follows (see also our introduction): On the one hand the notion of a free particle 
seems basic for the interpretation of ``mass'' and ``charge'' of an elementary 
particle. Also, in the case of perturbation theory this notion is crucial. On the other 
hand, the notion of ``freeness'' in this context refers to ``non-interaction'' and 
thus seems to contradict the dogma of gauge independence. However, the 
geometrical description of spontaneously broken gauge theories presented here 
permits to also describe the notion of a ``free particle'' in purely geometrical terms. 
Thus, the spontaneous symmetry breaking of the electroweak interaction permits 
the geometrical combination of the notion of a ``free particle'' with its ``mass'' and 
its ``charge''.\\ 

For this let $t\in[0,1]$. A one-parameter family of Yang-Mills-Higgs pairs 
$({\cal A}_t,\Phi_t)\in
{\cal A}(\xi_{\mbox{\tiny H}})\times\Gamma(\xi_{\mbox{\tiny H}})$ is called a
``linear fluctuation'' of a bosonic vacuum (pair) $(\Theta,{\cal V})$ provided that
\bb
\label{linearfluc}
{\cal A}_t &=& \Theta + t({\cal A} - \Theta),\cr
\Phi_t &=& {\cal V} + t\Phi.
\ee
Here, ${\cal A}\in{\cal A}(\xi_{\mbox{\tiny H}})$ is a connection associated 
with a  principal connection on ${\cal P}$ which is non-compatible 
with the vacuum; $\Phi\in\Gamma(\xi_{\mbox{\tiny H}})$ is a section
in the unitary gauge\footnote{Despite the terminology used the unitary ``gauge''
is in fact not a choice of gauge. Instead it refers to a specific choice of vacuum 
section adapted to the section $\Phi$ under consideration (see \cite{tolk1'03}).}, 
i. e. $\Phi$ uniquely corresponds to a section 
$\Phi_{\mbox{\tiny H,phys}}\in\Gamma(\xi_{\mbox{\tiny H,phys}})$. Of course,
the definition (\ref{linearfluc}) is the geometrical analogy to (\ref{mincoup}).\\

With respect to $(\Theta,{\cal V})$ the connection ${\cal A}$ uniquely corresponds 
to a section $A\in\Gamma(\xi_{\mbox{\tiny YM}})$. Moreover, because of the 
decomposition (\ref{ymg}) we have 
\bb
\label{asymfree}
A = A_{\mbox{\tiny elm}} + Z^0 + W^\pm,
\ee 
where
$A_{\mbox{\tiny elm}}\in\Gamma(\xi_{\mbox{\tiny elm}})\simeq\Omega^1(\mm),\,
Z^0\in\Gamma(\xi_{\mbox{\tiny${\rm Z}^0$}})\simeq\Omega^1(\mm)$ and
$W^\pm\in\Gamma(\xi_{\mbox{\tiny${\rm W}^\pm$}})\simeq\Omega^1(\mm)$.\\
 
The Euler-Lagrange equations of the Yang-Mills-Higgs functional 
${\cal I}_{\mbox{\tiny YMH}}$ with respect to the fluctuation 
$({\cal A}_t,\Phi_t)$ up to ${\cal O}(t)$ yield the well-known 
``free field equations''
\bb
\label{freefieldeq}
\delta\!d A_{\mbox{\tiny elm}} &=& 0,\cr
\delta\!d Z^0 + {\rm m}_{\mbox{\tiny Z}}^2 Z^0 &=& 0,\cr
\delta\!d W^\pm + {\rm m}_{\mbox{\tiny W}}^2 W^\pm &=& 0,\cr
\delta\!d\Phi_{\mbox{\tiny H,phys}} + 
{\rm m}_{\mbox{\tiny H,phy}}^2\Phi_{\mbox{\tiny H,phys}} &=&0.
\ee
Note that these second order equations are indeed H-invariant 
(but not G-invariant) and that 
$\delta Z^0=\delta W^\pm = \delta\Phi_{\mbox{\tiny H,phys}} = 0$ for reasons 
of consistency. Here, ``$d$'' denotes the exterior covariant derivative with
respect to the trivial connection, ``$\delta$'' its covariant co-derivative and
${\rm m}_{\mbox{\tiny H,phy}}\in{\rm spec}({\cal V}^*{\rm M}^2_{\mbox{\tiny H}})$ 
the ``mass'' of the physical Higgs boson\footnote{In the case of rotationally
symmetric Higgs potentials the rank of the physical Higgs bundle equals one, 
which equals the rank of the Higgs mass matrix.}, where 
${\cal V}^*{\rm M}^2_{\mbox{\tiny H}}\in
\Gamma({\rm End}({\rm E}_{\mbox{\tiny H}}))$ is the Higgs mass matrix
(c.f. \cite{tolk1'03}). Also note that
\bb
\delta\!d W^+ + {\rm m}_{\mbox{\tiny W}}^2 W^+ =
{\cal J}\left(\delta\!d W^- + {\rm m}_{\mbox{\tiny W}}^2 W^-\right).
\ee

Since, with respect to any geodesic coordinate system (local ``inertial system''), 
the principal symbols $\sigma_{\!\mbox{\tiny pr}}$ of the above second order 
differential operators asymptotically ($t\rightarrow 0$) equals the total symbols 
$\sigma,$ one may consider the above sections (\ref{asymfree}) and
$\Phi=\Phi_{\mbox{\tiny H,phys}}$ as geometrically representing states 
of asymptotically free particles (in semi-classical approximation)
that are represented by the corresponding (trivial) line bundles. Let 
$\Box:=\delta d + d\delta$ be the covariant wave operator (``d'Alambert operator'') 
with respect to the canonical connections on the trivial line bundles 
representing, respectively, the photon $\xi_{\mbox{\tiny elm}}$ the 
massive and (un-)charged vector bosons of the weak interaction,
$\xi_{\mbox{\tiny${\rm Z}^0$}},\,\xi_{\mbox{\tiny${\rm W}^\pm$}}$, and the 
physical Higgs boson $\xi_{\mbox{\tiny H,phys}}$. Also let
${\cal C}^+(\mm)\subset{\rm T}^*\mm$ be (pointwise) the future oriented part 
of the light cone that is defined by the Lorentz structure $g_{\mbox{\tiny M}}$. 
Then,  for all $\xi\in{\cal C}^+(\mm)$ the free field equations (\ref{freefieldeq}) 
indeed imply the well-known dispersion relation between mass, energy and 
momentum of a non-interacting pointlike particle
\bb
\label{dispersionrel}
\sigma(-\Box)(\xi) \stackrel{\ast}{=} \sigma_{\!\mbox{\tiny pr}}(-\Box)(\xi) =
g_{\mbox{\tiny M}}(\xi,\xi) = \left\{\begin{array}{r@{\quad:\quad}l}
0 & \xi_{\mbox{\tiny elm}}\\
{\rm m}_{\mbox{\tiny${\rm Z}^0$}}^2,\,{\rm m}_{\mbox{\tiny W}}^2 & 
\xi_{\mbox{\tiny${\rm Z}^0$}},\,\xi_{\mbox{\tiny${\rm W}^\pm$}}\\
{\rm m}_{\mbox{\tiny H,phys}}^2 & \xi_{\mbox{\tiny H,phys}}
\end{array}\right.
\ee
Here, ``$\ast$'' means ``with respect to any geodesic coordinate system''.\\

While the local form of (\ref{freefieldeq}) and (\ref{dispersionrel}) can be
found in almost any text book of quantum field theory we summarized them
here to put emphasis on their geometrical content. In fact, the first equality
of the relations (\ref{dispersionrel}) might serve as a geometrical definition of
``freeness'', the second equality combines the notions of particle and of field
and the third equality indicates the geometrical background of the former
two equalities in the realm of (spontaneously broken) gauge theories. Notice
that the right hand side of the dispersion relation (\ref{dispersionrel}) is fully
determined by the sections ${\cal V}^*{\rm M}^2_{\mbox{\tiny YM}}\in
\Gamma({\rm End}({\rm ad}({\rm P}))),\,{\cal V}^*{\rm M}^2_{\mbox{\tiny H}}\in
\Gamma({\rm End}({\rm E}_{\mbox{\tiny H}})$ together with the assumption
that electromagnetism is invariant with respect to charge conjugation. That
the spectrum of these sections is well-defined (i. e. constant and independent of 
the vacuum section chosen) is part of the mathematical structure of 
spontaneously broken gauge theories. The appropriate physical interpretation 
of the spectrum, however, is motivated only by the first two equalities on the 
left hand side of (\ref{dispersionrel}).\\

We close this section with a discussion on the geometrical meaning of 
the physical particles (\ref{physparticle}) of the electroweak interaction.
For this we consider the Higgs bundle $\xi_{\mbox{\tiny H}}$ as a real 
vector bundle of rank four. Each YMH pair $({\cal A},\Phi)$ yields a 
specific embedding of $\mm$ into the total space ${\rm E}_{\mbox{\tiny H}}$ 
of the Higgs bundle and a specific splitting of the tangent bundle 
$\tau_{\mbox{\tiny H}}$ of ${\rm E}_{\mbox{\tiny H}}$ into its horizontal 
and vertical part
\bb
\label{decom}
\tau_{\mbox{\tiny H}}=
{\goth{h}}_{\mbox{\tiny H}}\op{\goth{v}}_{\mbox{\tiny H}}.
\ee
Moreover, $(g_{\mbox{\tiny M}},{\cal A})$ turns ${\rm E}_{\mbox{\tiny H}}$
into a (semi-)Riemannian manifold of dimension $4{\rm dim}\mm$ such
that the splitting (\ref{decom}) becomes orthogonal.\\

Let $(\Theta,{\cal V})$ be an electroweak vacuum and $({\cal A},\Phi)$ 
be a (linear) fluctuation thereof. We call 
$\mm_{\mbox{\tiny phys}}:={\cal V}(\mm)$ the ``physical space-time'' with 
respect to the vacuum $(\Theta,{\cal V})$. Accordingly, we denote by 
$g_{\mbox{\tiny H}}$ the (pseudo) metric on ${\rm E}_{\mbox{\tiny H}}$ with 
respect to $(g_{\mbox{\tiny M}},\Theta)$. Because of
$\xi_{\mbox{\tiny orb}}\subset\xi_{\mbox{\tiny H}}$ one obtains the following 
orthogonal decompositions along $\mm_{\mbox{\tiny phys}}$:
\bb
\label{geophysparticlea}
\tau_{\mbox{\tiny H}}|_{\mbox{\tiny M,phys}}&\simeq&
\tau_{\mbox{\tiny Orb}}|_{\mbox{\tiny M,phys}}\op
\nu_{\mbox{\tiny Orb}}|_{\mbox{\tiny M,phys}}\nonumber\\[0.2cm]
&\simeq&
{\goth{h}}_{\mbox{\tiny Orb}}|_{\mbox{\tiny M,phys}}\op
\pi^*_{\mbox{\tiny H}}\xi_{\mbox{\tiny G}}|_{\mbox{\tiny M,phys}}\op
\pi^*_{\mbox{\tiny H}}\xi_{\mbox{\tiny H,phys}}|_{\mbox{\tiny M,phys}}\nonumber\\[0.2cm]
&\simeq&
\tau_{\mbox{\tiny M,phys}}\op
\pi^*_{\mbox{\tiny H}}\xi_{\mbox{\tiny Z}}|_{\mbox{\tiny M,phys}}\op
\pi^*_{\mbox{\tiny H}}\xi_{\mbox{\tiny${\rm W}_1$}}|_{\mbox{\tiny M,phys}}\op
\pi^*_{\mbox{\tiny H}}\xi_{\mbox{\tiny${\rm W}_2$}}|_{\mbox{\tiny M,phys}}\op
\pi^*_{\mbox{\tiny H}}\xi_{\mbox{\tiny H,phys}}|_{\mbox{\tiny M,phys}}.
\ee
Here, respectively, $\tau_{\mbox{\tiny Orb}} and \nu_{\mbox{\tiny Orb}}$ denote
the tangent and the normal bundle of the orbit bundle and
$\tau_{\mbox{\tiny M,phys}}$ is the tangent bundle of the physical
space-time.\\

Correspondingly, for every 
$w=({\cal V}(x),{\bf w})\in{\rm T}{\rm E}_{\mbox{\tiny H}}$ there are real 
constants $\lambda_{\mbox{\tiny elm}}, \lambda_{\mbox{\tiny 0}},$
$\lambda_{\mbox{\tiny 1}}, \lambda_{\mbox{\tiny 2}}$ and
$\lambda_{\mbox{\tiny H, phys}}$ such that $w$ can be 
written in terms of the solutions of the free field equations 
(\ref{freefieldeq}):
\bb
\label{geophysparticleb}
w = \lambda_{\mbox{\tiny elm}}\,{\rm T}({\cal V}\circ\pi_{\mbox{\tiny H}})(w) + 
\lambda_{\mbox{\tiny 0}}\, {\bf Z}^0(w) + \lambda_{\mbox{\tiny 1}}\, {\bf W}^1(w) + 
\lambda_{\mbox{\tiny 2}}\, {\bf W}^2(w) + 
\lambda_{\mbox{\tiny H, phys}}\,\Phi_{\mbox{\tiny H,phys}}(x).
\ee
Here, for instance, ${\bf Z}^0(w)\equiv\rho'_{\mbox{\tiny H}}
(\pi^*_{\mbox{\tiny H}}Z^0(w)){\cal V}\in
{\rm T}{\rm E}_{\mbox{\tiny H}}|_{\mbox{\tiny M,phys}}$ etc.,
and $\rho'_{\mbox{\tiny H}}:={\rm d}\rho_{\mbox{\tiny H}}(e)$ is the (real form 
of the) induced representation on Lie(G). Notice that we have made use of the
Higgs dinner (\ref{higgsdinner}) in such a way that $\pi^*_{\mbox{\tiny H}}Z^0\in
\Omega^1({\rm E}_{\mbox{\tiny H}},{\rm ad(P)})$ and that the gauge group
${\cal G}$ of the electroweak gauge bundle naturally acts from the right on
the orbit bundle.\\

As was shown in \cite{tolk1'03} the restriction to $\mm_{\mbox{\tiny phys}}$ 
of any compatible connection ${\cal A}_{\mbox{\tiny elm}}$ on $\xi_{\mbox{\tiny H}}$
coincides with the canonical ``connection'' that is defined by the vacuum section 
${\cal V}$. That is, if we denote by $\wp^{\mbox{\tiny H}}_{\!\mbox{\tiny elm}}$ the
horizontal projector on $\tau_{\mbox{\tiny H}}$ with respect to the
connection ${\cal A}_{\mbox{\tiny elm}}$ then
\bb
\label{canlift}
\wp^{\mbox{\tiny H}}_{\!\mbox{\tiny elm}}|_{\mbox{\tiny${\cal V}(x)$}}({\bf w})=
{\rm d}{\cal V}(x)({\rm d}\pi_{\mbox{\tiny H}}({\cal V}(x)){\bf w}),
\ee
for all $x\in\mm$ and ${\bf w}\in
{\rm T}_{\mbox{\tiny${\cal V}(x)$}}{\rm E}_{\mbox{\tiny H}}$. Moreover,
with respect to the decomposition (\ref{asymfree}) any connection ${\cal A}$ 
on the Higgs bundle can be written as
\bb
{\cal A}  &:=& \Theta + {\bf A}\cr
&=& {\cal A}_{\mbox{\tiny elm}} + {\bf A}_{\mbox{\tiny G}}.
\ee
Here, ${\cal A}_{\mbox{\tiny elm}} = 
\Theta + {\bf A}_{\mbox{\tiny elm}}$ and
${\bf A}_{\mbox{\tiny G}}\equiv {\bf Z}^0 + {\bf W}^1 + {\bf W}^2$. Then, the 
horizontal projector with respect to an arbitrary (associated) connection 
${\cal A}\in\Gamma(\xi_{\mbox{\tiny H}})$ reads
\bb
\label{hproj}
\wp^{\mbox{\tiny H}}_{\!\mbox{\tiny A}} =
\wp^{\mbox{\tiny H}}_{\!\mbox{\tiny elm}} + {\bf A}_{\mbox{\tiny G}}
\ee 
and thus
\bb
\wp^{\mbox{\tiny H}}_{\!\mbox{\tiny A}}|_{\mbox{\tiny M,phys}}(w)=
{\rm T}({\cal V}\circ\pi_{\mbox{\tiny H}})(w) + {\bf A}_{\mbox{\tiny G}}(w)
\ee
for all $w=({\cal V}(x),{\bf w})\in{\rm T}{\rm E}_{\mbox{\tiny H}}$.\\

Since the decomposition (\ref{geophysparticlea}) is orthogonal with respect
to $g_{\mbox{\tiny H}}$ one obtains for all 
$w_i=({\cal V}(x),{\bf w}_i)\in{\rm T}{\rm E}_{\mbox{\tiny H}}\,$ $(i=1,2)$
\bb
\label{higgsmetric}
g^{\mbox{\tiny A}}_{\mbox{\tiny H}}|_{\mbox{\tiny M,phys}}(w_1,w_2) &:=&
\kappa_{\mbox{\tiny H}}
(\wp^{\mbox{\tiny V}}_{\!\mbox{\tiny A}}|_{\mbox{\tiny M,phys}}(w_1),
\wp^{\mbox{\tiny V}}_{\!\mbox{\tiny A}}|_{\mbox{\tiny M,phys}}(w_2)) +
\pi^*_{\mbox{\tiny H}}g_{\mbox{\tiny M}}(w_1,w_2)\nonumber\\[0.15cm]
&=&
g_{\mbox{\tiny H}}|_{\mbox{\tiny M,phys}}(w_1,w_2) +
\kappa_{\mbox{\tiny H}}({\bf A}_{\mbox{\tiny G}}(w_1),
{\bf A}_{\mbox{\tiny G}}(w_2)),
\ee
where, respectively, $\wp^{\mbox{\tiny V}}_{\!\mbox{\tiny A}}$ is the vertical 
projector that is defined by (\ref{hproj}) and $\kappa_{\mbox{\tiny H}}$ is the 
Hermitian form on the Higgs bundle.\\

The relation (\ref{higgsmetric}) shows that the massive weak vector bosons 
$(Z^0,W^\pm)$ correspond to normal sections of $\mm_{\mbox{\tiny phys}}$
which yield a ``fluctuation'' of the (pseudo) metric $g_{\mbox{\tiny H}}$. In contrast, 
the massless photon $A_{\mbox{\tiny elm}}$ only gives rise to a change of 
$g_{\mbox{\tiny H}}$ off the physical space-time $\mm_{\mbox{\tiny phys}}$.
Note that, when restricted to $\mm_{\mbox{\tiny phys}},$ the connection
${\cal A}_{\mbox{\tiny elm}}$ is flat. In particular, one obtains 
$d_{\!\mbox{\tiny A,elm}}{\cal V} = 0$ and thus 
\bb
\label{convac}
d_{\!\mbox{\tiny A}}{\cal V} = {\bf A}_{\mbox{\tiny G}}. 
\ee
We stress that it is this relation between an arbitrary connection on the
electroweak gauge bundle ${\cal P}$ and the electroweak vacuum that yields 
a non-trivial Yang-Mills mass matrix (\ref{ymm}).\\

\section{Remarks} 
In the following we give some comments on the results presented.\\

The presented classification theorem concerning the bosonic vacua in the 
case of the electroweak interaction can be generalized to more general 
YMH data $({\cal P},\rho_{\mbox{\tiny H}},{\rm V}_{\!\mbox{\tiny H}})$
whereby the ``little group'' H is supposed to be given either by U(1), or by SU(2). 
The general Higgs potential is, again, assumed to be rotationally symmetric.\\

In the slightly more general case with H = U(1), the moduli space of bosonic vacua 
can be identified with 
\bb
{\goth{M}}_{\mbox{\tiny vac}} = 
{\rm H}^1_{\mbox{\tiny deR}}(\mm)\times\#{\rm Orb},
\ee 
where $\#{\rm Orb}$ denotes the number of orbits.\\ 

For instance, in the case of a ``Sine-Gordon''-like Higgs potential 
\bb
{\rm V}_{\!\mbox{\tiny H}}({\bf z}):=\mbox{\small$\frac{\mu^2}{\lambda^2}$}
\left(1-\cos(\lambda|{\bf z}|)\right)\qquad (\mu,\lambda >0),
\ee 
the moduli space of bosonic vacua reads 
\bb
{\goth{M}}_{\mbox{\tiny vac}} = {\rm H}^1_{\mbox{\tiny deR}}(\mm)\times\zz.
\ee 

In the case where the isotropy group of a minimum ${\bf z}_0$ of a general Higgs 
potential can be identified with SU(2) one obtains
\bb
{\goth{M}}_{\mbox{\tiny vac}} = \goth{M}\times\#{\rm Orb},
\ee
where $\goth{M}$ is the moduli space of flat SU(2)-connections on the 
trivial principal SU(2)-bundle over space-time $\mm$.\\

The results presented imply that the set of non-vanishing sections of the 
Higgs bundle is in one-to-one correspondence with the non-vanishing smooth 
real-valued functions on space-time $\mm,$ i. e.
\bb
\label{higgsfunc}
\Gamma^*(\xi_{\mbox{\tiny H}})\equiv
\Gamma(\xi_{\mbox{\tiny H}})\backslash\{{\cal O}\}\simeq
{\cal C}^\infty(\mm)\backslash\{0\},
\ee 
where ${\cal O}$ (resp. $0$) is the zero section (resp. zero function) on $\mm$.
From theorem (\ref{satz}) it follows that the Higgs bundle is trivial and therefore
$\Gamma(\xi_{\mbox{\tiny H}})\simeq\Omega^0(\mm,\cc^2).$ If one identifies
$\cc^2\simeq\hhh$ with the quaternions via ${\bf z} = (z_1,z_2)\mapsto
q = z_1 + z_2 j,$ then one may make use of the polar decomposition for non-zero
quaternions $q = ||q||\exp(\vartheta\, n)$ ($n\in\hhh,\; n^2 = -1$) to show that
$\Phi\in\Gamma^*(\xi_{\mbox{\tiny H}})$ is gauge equivalent to the mapping
\bb
\label{standardhiggs}
\mm&\longrightarrow&{\rm E}_{\mbox{\tiny H}}\cr
x&\mapsto& (x, ||\Phi(x)||{\bf e}_0)
\ee
with ${\bf e}_0:={\bf z}_0/||{\bf z}_0||\in{\rm S}^3$.\\ 

Accordingly, a section of the physical Higgs bundle $\xi_{\mbox{\tiny H,phys}}$ 
reads
\bb
\label{unigauge}
\Phi_{\mbox{\tiny H,phys}}:\,\mm&\longrightarrow&
{\rm E}_{\mbox{\tiny H,phys}}\cr
x&\mapsto& (x, ||\Phi(x)||),
\ee 
where we have identified 
$\rr^4\supset{\rm W}_{\mbox{\tiny H,phys}}:=\rr\,{\bf e}_0$ with $\rr$.
Moreover, with help of this identification the free field equation for the physical 
Higgs boson is reduced to the ordinary Klein-Gordon equation for the function 
$\varphi:=||\Phi||$. \\

The mapping (\ref{standardhiggs}) is known in physics as the ``Higgs boson in the
unitary gauge''. As is well-known, this terminology refers to the fact that the
``phase'' of a particle is unphysical and can be thus ``gauged away''. In fact, when
considered as a field, $n\in\rr^3\subset\hhh$ geometrically corresponds to a 
section of the Goldstone bundle $\xi_{\mbox{\tiny G}}.$ For the above given
argument which is used in physics to show the existence of the unitary gauge
in the electroweak interaction and which leads to (\ref{higgsfunc}) it seems crucial 
that the ``phase of the Higgs boson'' can be identified with an element of 
${\rm SU(2)}\subset{\rm G}.$ However, this turns out not to be the case, actually.
Indeed, the isomorphism (\ref{higgsfunc}) only depends on the structure of the
Higgs potential and not, e. g., of the simple structure of 
${\goth{M}}_{\mbox{\tiny vac}}.$ In particular, the existence of the unitary gauge
does not depend on the triviality of the Higgs bundle. For example, in contrast to 
the above given argument (which only works in the case where ${\cal P}$ is trivial), 
the isomorphism (\ref{higgsfunc}) always holds true for rotationally symmetric 
Higgs potentials. In this case, the physical Higgs bundle must be necessarily
trivial for every vacuum section and thus also (\ref{unigauge}) generally holds 
true (c. f. \cite{tolk1'03}). For rotationally symmetric Higgs potentials the question 
about the existence of the unitary gauge is related to the question whether
$\Gamma^*(\xi_{\mbox{\tiny H}})$ is empty or not. But this is basically the same as 
to ask about the existence of vacuum sections which spontaneously break the 
gauge symmetry. Therefore, the assumption that 
$\Gamma^*(\xi_{\mbox{\tiny H}})\not=\emptyset$ is physically well-motivated.
The results presented here with respect to the electroweak interaction show 
that in the case of ordinary electromagnetism 
$\Gamma^*(\xi_{\mbox{\tiny H}})\not=\emptyset$ is the same as the triviality 
of the electroweak gauge bundle (see proposition (\ref{prop})).\\ 

Although, in general, the notion of ``free particles'' depends on the gauge class 
$[(\Theta,{\cal V})]\in{\goth{M}}_{\mbox{\tiny vac}}$ of electroweak vacua, the 
geometrical interpretation of the particle content of the electroweak interaction
holds true for all vacua. Moreover, the relation (\ref{convac}) remains intact
for arbitrary connections ${\cal A}_{\mbox{\tiny elm}}$. It only makes use of
the compatibility condition (\ref{hred}) and the triviality (\ref{canlift}) of
${\cal A}_{\mbox{\tiny elm}}$ along the physical space-time 
${\cal M}_{\mbox{\tiny phys}}$. Notice that the latter geometrical property of
an H-reducible connection guarantees that the spectrum of the bosonic mass 
matrices is constant. Accordingly, the Higgs dinner (\ref{higgsdinner}) 
also works in the case when the gauge bundle ${\cal P}$ possesses no flat 
connections. This remark becomes important, for example, when non-trivial 
U(1)-reductions (resp. SU(2)-reductions) of ${\cal P}$ are considered. We 
close this section with the remark that the intrinsic geometry of the physical 
space-time ${\cal M}_{\mbox{\tiny phys}}$ is the same as that of the ``naked'' 
space-time $\mm$. We summarize this by saying that in the case of ordinary
electromagnetism the structure of the moduli space of the electroweak vacua 
${\goth{M}}_{\mbox{\tiny vac}}$ only depends on the topology of space-time 
but not on its geometry. This, however, may change for non-trivial vacua.\\

\section{Conclusion}
In this paper we have discussed the moduli space of bosonic vacua 
of the electroweak interaction on ``tree level''. We have proved that the 
corresponding moduli space is either empty or an affine space that can 
be canonically identified with the first de Rham cohomology group of 
space-time. We have shown that, when charge conjugation is taken into 
account, the existence of non-trivial ground states of the Higgs boson is 
equivalent to the triviality of the electroweak gauge bundle. For this, 
however, it is crucial that spontaneous symmetry breaking of the 
electroweak interaction not only yields massive but also electrically 
charged bosons. It follows that the electromagnetic gauge bundle must 
be also trivial. For this reason one may ask about the existence of magnetic 
monopoles within the realm of the Standard Model. Basically, there are 
two answers to this question: On the one hand, one may consider a magnetic 
monopole as a physical object in its own which is independent of the 
Standard Model. This, however, raises the question of the physical meaning 
of the underlying U(1) gauge bundle of the monopole and its relation to Dirac's 
quantization condition of electric charge. On the
other hand, since a monopole is assumed to be massive and related
to electromagnetism it seems far more natural to consider it as a possibly 
non-trivial electromagnetic reduction of the electroweak interaction. In this case,
a monopole is not considered as a separate particle but as a certain ground
state of the Higgs boson which is gauge inequivalent to the ground state 
usually encountered in perturbation theory (i. e. to ${\cal V}_{\!\mbox{\tiny 0}}$). 
In any case, the existence of a monopole field would spoil the symmetry under 
charge conjugation. Since the latter has been shown in this paper to be intimately 
related to the structure of the moduli space of electroweak vacua, it is natural to 
complete our discussion on the geometrical structure of the electroweak 
interaction by considering the case where charge conjugation ${\cal J}$ is not 
assumed to exist. This will be done in a forthcoming paper.

\vspace{1.5cm}

\noindent
{\bf Acknowledgments}\\
I am very grateful to E. Binz for stimulating discussions on 
the subject matter discussed in this paper.

\end{document}